\documentclass[prb,twocolumn]{revtex4}
\usepackage{graphicx,latexsym}

\begin{document}
\title{On the ground states of an array of magnetic dots in the vortex state and
subject to a normal magnetic field.}
\author{J.E.L.Bishop }
\affiliation{Department of Physics and Astronomy, The University of Sheffield,
Sheffield S3 7RH, U.K.}
\author{A.Yu.Galkin}
\affiliation{Institute of Metal Physics, National Academy of Sciences of Ukraine,
Vernadskii av. 36, 03142, Kiev, Ukraine }
\author{B.A.Ivanov}
\email{bivanov@i.com.ua}
\affiliation{Institute of Magnetism, National Academy of Sciences of Ukraine,
Vernadskii av. 36 ''B'', 03142, Kiev, Ukraine }
\date{\today }

\begin{abstract}
Dipole-dipole interactions in a square planar array of sub-micron magnetic
disks (magnetic dots) have been studied theoretically. Under a normal
magnetic field the ground-state of the array undergoes many structural
transitions between the limiting chessboard antiferromagnetic state at zero
field and the ferromagnet at a threshold field. At intermediate fields,
numerous ferrimagnetic states having mean magnetic moments between zero and
that of the ferromagnetic state are favorable energetically. The structures
and energies of a selection of states are calculated and plotted, as are the
fields required to optimally reverse the magnetic moment of a single dot
within them. Approximate formulae for the dipolar energy and anhysteretic
magnetization curve are presented.
\end{abstract}

\maketitle

\section{ Introduction}

\label{sec:introduction}

Recently some peculiar properties of sub-micron magnetic particles
(magnetic dots) fabricated from such soft magnetic materials as
permalloy, Co, etc., and forming an artificial lattice have
attracted great attention, see Refs. \onlinecite
{Hillebrand1}-\onlinecite{Cowburn5}. These magnetic dot arrays
constitute promising material for high-density magnetic storage
media. The distribution of magnetization within the dots is quite
nontrivial. In the absence of an external magnetic field, a small
enough non-ellipsoidal dot exhibits a single-domain nearly uniform
magnetization state, either a so called flower state or a leaf
state.\cite{Usov6} On increasing the size of the dot above a
critical value, a vortex state occurs.\cite{Usov7} This vortex
state has been experimentally observed ( see Refs.
\onlinecite{Runge4}, \onlinecite{Cowburn5}, \onlinecite
{Fernandez8} - \onlinecite{Pokhil10}) for circular disk-shaped
magnetic dots with diameters $2R=200-800$ nm and thickness
$L=20-60$ nm. In Ref. \onlinecite{Novosad11}, magnetization
reversal for an array of disk-shaped dots under the influence of a
magnetic field applied in the plane of the dots has been
investigated experimentally. In this planar geometry the main
contribution to the total magnetization comes from the internal
reorganization of each dot's magnetic structure, in particular, by
displacement of the vortex from the dot center, leading eventually
to annihilation of the vortex at the rim of the dot. In this
process, dipolar interaction between the dots does not play an
essential role.

In the present work, another case, namely, the ground state of an
unbounded planar square lattice of thin circular disk-shaped dots
in an external magnetic field perpendicular to the plane of the
dots will be considered. We will show that the situation in this
case is very different from that in which the field is applied
in-plane: the main contribution to the total magnetization is now
determined as much by the dipolar interactions of the dots as by
the external field. In a perpendicular field, the dipole-dipole
interaction between the dots results in a complex specific phase
diagram. In particular, a cascade of phases with different
patterns of dot magnetization has been found; these constitute the
sequence of ground states as a function of the external magnetic
field.

\section{Model}

\label{sec:model}

In order to formulate the model we need to discuss briefly the
character of $ \vec M$, the magnetization distribution within a
single dot in the vortex state. In circular cylindrical
coordinates, $\vec {M} =\vec M(z,r,\chi )$. In sufficiently thin
material, such as that being considered in the present work, $\vec
{M}$ does not depend significantly on the $z$-coordinate along the
normal to the dot, so that we may more simply write $ \vec
{M}=\vec {M}(r,\chi )$, where $r$ and $\chi $ are the polar
coordinates in the dot plane. Then the Cartesian components of
$\vec {M}$ for the vortex state inside the dot, $M_x=M_s\sin
\theta \cos \varphi ,  M_y=M_s\sin \theta \sin \varphi $,
$M_z=M_s\cos \theta $, where $M_s$ is the saturation
magnetization, are determined by ansatz

\begin{equation}
\theta =\theta (r),\varphi =\chi +\varphi _0.  \label{ansatz}
\end{equation}

Such a distribution is typical for magnetic vortices with a
topological charge (vorticity) equal to one, in two-dimensional
easy-plane ferromagnets. The function $\theta (r)$ is determined
by an ordinary differential equation and its solution can be
easily found numerically -- see the general discussion of magnetic
vortices in Refs. \onlinecite{Kosevich12}, \onlinecite{Ivanov13}.

For the theory of magnetic dots made of soft magnetic materials,
the crystallographic easy-plane anisotropy is negligible, and the
demagnetizing field $\vec {H}_m$ plays the main role. The sources
of $\vec {H}_m$ are both the volume "magnetic charges",
proportional to $\text{div}\vec {M}$, and the discontinuity in the
normal component of magnetization at the surface of the sample
(the surface "magnetic charges"). The vortex distribution
(\ref{ansatz}) has an advantage compared with others, because for
(\ref{ansatz}) $\text{div}\vec {M}=M_s\cos \varphi _0\left(
{\displaystyle {d\theta  \over dr}} \cos \theta + {\displaystyle
{1 \over r}} \sin \theta \right) $, and $\text{div}\vec {M}=0$ at
$\varphi _0=\pm \pi /2$. Accordingly, the volume magnetic charges
vanish and the sole source of the field is the $z-$projection of
$\vec {H}_m$ onto the faces of the dot. The states with $\varphi
_0=+\pi /2$ and $\varphi _0=-\pi /2$ have the same energy, i.e.
the vortex state of the dot is twofold degenerate with respect to
the sense of the magnetization rotation. In the case of a thin
enough dot, this gives $\vec {H}_m=-4\pi M_s\hat{\vec{z}},$ i.e.
there is an effective easy plane anisotropy $w_m=2\pi M_s^2\cos
^2\theta $.

The function $\theta (r)$ can be obtained by applying well-known
methods for treating magnetic vortices in easy plane magnets. In
the center of the dot ($ r=0$) the function $\theta (r)$ is
restricted to two possible values $\theta (r=0)=0,\pi $ ; so that
$\cos \theta =p=\pm 1$. Here $p$ is the so-called vortex
polarization (the second topological charge).\cite{Ivanov13} The
characteristic scale of the variation of the function $\theta (r)$
coincides with the value of the exchange length $\Delta _0$
\begin{equation}
\Delta _0=\sqrt{A/4\pi M_s^2},  \label{exchangelen}
\end{equation}
where $A$ is the inhomogeneous exchange constant.

For $r\gg \Delta _0$ in the vortex solution, the value $\theta
(r)$ tends to $\pi /2$ exponentially. If the dot radius
$R\gg\Delta _0$, then one can obtain an acceptable solution in
which the limiting value $\theta (r)=\pi /2$ is reached at $r=R$,
the rim of the dot. In this case $\theta \approx \pi /2$ within
the major part of the volume of the dot, where $\Delta _0$
\mbox{$<$} $r\leq R$. Such states have been discussed in a
theoretical treatment of magnetic dots that included an exact
treatment of the magnetic dipole-dipole interaction.\cite{Usov7}
That treatment showed that the out-of-plane magnetization $M_z$ is
significantly different from zero only in the core region, $r\leq
\Delta _0$, and that the total magnetic moment of the dot $ \vec
{\mu }$ has one of two values:

\begin{equation}
\vec {\mu }=p\mu \hat {\vec{z}},p=\pm 1,\mu =2\pi \xi L\Delta
_0^2M_s,  \label{magnetiz}
\end{equation}
where $\xi $ is a multiplicative constant of the order of 1; in
fact $\xi  \rightarrow \ 1.361$ as $\Delta _0/R \rightarrow $\ 0.

Thus, we arrive at the following simplified picture of the state
of the single dot. In the greater part of the dot, the
magnetization lies in the dot plane, rotating about the center of
the dot, and so, because of its circular symmetry, does not
contribute to the total moment $\vec {\mu }$ of the dot.\cite
{Usov7} The state of the dot is fourfold degenerate, with $\varphi
_0=\pm \pi /2$ and the core polarization $p=\pm 1$. The value of
$\varphi _0$ does not manifest itself in the magnetic moment of
the dot and, as a consequence, has no influence on dot
interactions. The magnetic moment is directed perpendicular to the
plane of the dot and has either of the values $\pm \mu \hat
{\vec{z}}$.

With regard to dot interactions, all dots are in one or other of
two states: "up" and "down". Since the core volume is much smaller
than the dot volume, the core magnetic moment is small compared
with the saturation moment of the dot. Although the dipolar
interaction between dots is not very strong,\cite{Guslienko14} it
is nevertheless the sole source of interaction within the dot
system. Because both the dipolar dot interaction field and the
external magnetic field are very much lower than the effective
fields (exchange and demagnetizing) internal to the dot, one can
regard the magnetization distribution inside the dot as
practically unaffected by them.

An important consequence of this robustness of the vortex state is
that the polarization $p$ and moment $\vec {\mu }$ of a dot remain
unchanged under the application of small enough external magnetic
fields $\vec {H}_e$ parallel to $\hat {\vec{z}}$.\cite{Ivanov15}
It is clear that not only the state with $ \vec {H}_e\| \vec
{\mu}$ (light vortex) is stable, but also the state $\vec {H}_e$
with antiparallel to $ \hat {\vec{z}}$ (heavy vortex) remains
constant and metastable up to $ |\vec {H}_e|=4\pi
M_s$.\cite{Ivanov15}

Following from the above discussion, the Hamiltonian of a dot
array can be presented as follows:

\begin{equation}
{\mathcal{H}}= {\displaystyle {\mu ^2 \over 2a^3}}
\sum\limits_{\vec {l}\neq \vec {l^{^{\prime }}}} {\displaystyle
{p_{\vec {l}}p_{\vec {l^{^{\prime }}}} \over \left| \vec {l}-\vec
{l^{^{\prime }}}\right| ^3}} -\mu H\sum\limits_{\vec {l}}p_{\vec
{l}}, \label{dothamilt}
\end{equation}
where $\mu $ is the moment of a single dot, $p_{\vec {l}}=\pm 1,
\vec {l},\vec {l^{^{\prime }}}$ are dot positions in a square
lattice, $\vec {l}=a(m\hat{\vec {x}}+n\hat{\vec {y}}),  m,n=0,\pm
1,\pm 2$, ... are integers, $a$ is the interdot distance, and $H$
is the external magnetic field parallel to $ \hat {\vec{z}}$. The
first term describes the dipole-dipole interaction of the lattice
of magnetic dots, the second term is the Zeeman energy.

It should, perhaps, be stressed that the system modelled here,
consisting of a square lattice of discrete dipoles, normal to the
lattice plane, with only dipolar interactions, is very different
from a continuous thin film with perpendicular anisotropy.
Numerical treatments of such films, when performed using a square
lattice discretization, eg. see Refs. \onlinecite{Abanov16},
\onlinecite{Arlett17}, give rise to a Hamiltonian that bears a
superficial resemblance to that of the present system, but the
essential continuity of the film, expressed in the exchange
coupling between nearest neighbour elements of the numerical
discretization - the dominant interaction in any sufficiently
refined discretization - necessarily results in magnetization
patterns (stripe domain structures) that are entirely different
from the patterns of discrete moments arising from pure dipolar
interactions between discrete dots on a square lattice that are
reported here.

\section{MAGNETIC GROUND STATES}

\label{sec:magnetic states}

The dipole-dipole interaction is long-range, and it is not obvious
\textit{a priori} what structure will constitute the ground state.
For a system of particles with dipole-dipole interactions and in
zero applied field, a theorem states that in the ground state the
overall magnetic moment is zero, and this results in a specific
antiferromagnetic (AFM) state.\cite {Luttinger18} For instance, in
the case of a three-dimensional simple cubic lattice of spherical
particles, a four-sub-lattice structure with non-collinear
magnetic moments is optimal.\cite{Belobor19} This is possible,
however, only if the magnetic moment of each particle is free to
point in any direction. Here, because of the robustness of the
vortex state supporting the dot magnetic moment, we have $p=\pm 1$
and uniaxial, giving rise to a quasi-Ising model. Determination of
the ground state is reduced to geometrical considerations.

It is convenient to divide the initial simple square lattice, on
which the dots are located, into elementary \textit{magnetic}
cells of rectangular shape with $(k\times l)$ dots, so that the
overall spatial arrangement of up and down $(p=\pm 1)$ dots (what
we call their "structure", "configuration" or "pattern") can be
produced from a single such cell by a translation $\vec {T}=a(N
\hat {\vec {x}} +M\hat {\vec {y}})$, where $M=km$ and $N=ln$ and
$m,n=0,\pm 1,\pm 2$, ... are integers. This is appropriate for a
magnetic structure with sublattice number (i.e. smallest
rectangular unit cell size) less than or equal to $k\times l$.
Note that as the choice of specific values for $k$ and $l$
restricts the range of structures that can be represented in this
way, the search for the ground state must, in principle, extend to
all integer $k,l$. The Zeeman energy, however, depends only on the
relative numbers of dots with $p=+1$ and $p=-1$ in a magnetic
cell, i.e. on the mean moment per dot $\left\langle \mu
\right\rangle (=\left\langle \mu _z\right\rangle )$, and, of
course, on the applied field $H$.

\begin{figure}
\includegraphics[width=3.5in]{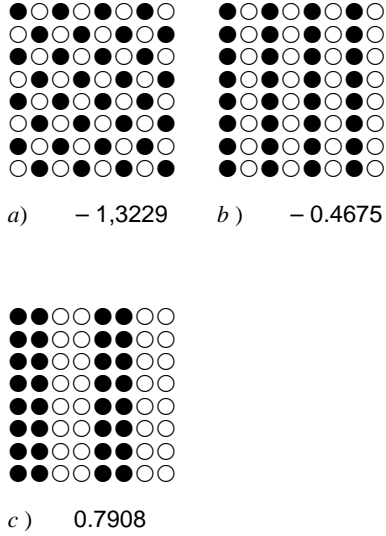}
\caption{ \label{Fig1}  Different antiferromagnetic states for a
square magnetic dot lattice. The dipolar energies per dot of the
lattices (in units of ($\mu ^2/a^3$ - see text) have been found
numerically and printed below each sketch. Open circles ${\circ}$
denote the "up" dots with moments $\vec \mu = \mu \hat{\vec z} $;
solid circles  ${\bullet}$  correspond to "down" dots with $\vec
\mu = -\mu \hat{\vec z} $.}
\end{figure}

We have investigated a substantial number of states, namely, all
states with $k\times l=2,3$ and $4$, and many states with larger
values of $k$ and $ l$. Because the dipolar interaction falls off
with the inverse cube of the moment separation, i.e. quite
rapidly, it is clear that distributions in which dots of like
orientation are well apart from each other, while those of
opposite sign are as close as possible, will be energetically most
favourable. In particular, one notes that the ratio of the
interaction energy of a nearest neighbour dot pair, to that of a
pair of next-nearest neighbours, is $\sqrt{8}:1$. It seems
extremely improbable that two dots of the minority population
(which we will consistently take to be the down dots with $p=-1$)
could ever be nearest neighbours in a ground-state configuration.
In zero magnetic field, the most energetically favorable
distribution is the simple chessboard AFM structure, see Fig.\
\ref{Fig1}. In this structure neither of the two (equal)
populations contains any nearest neighbour pairs of parallel dots.
Other AFM structures in which dots with the same value of $\vec
{\mu }$ do occur as nearest neighbours possess very much higher
energy as is illustrated by the examples in Fig.\ \ref{Fig1}.

For the AFM structure, the energy does not depend on the applied
magnetic field, whereas for the ferromagnetic (FM) structure (with
 $\vec\mu = +\mu\hat {\vec {z}}$ for all dots) this dependence is
maximal. The total mean energy per dot can be written as $W  = W_m
-\left\langle \mu \right\rangle H$, where $W_m$ is the mean
energy, per dot, of the dipole-dipole interaction and
$\left\langle \mu \right\rangle $ is the overall average moment
per dot of the distribution (zero in the AFM case). In virtue of
this, the energies $W_{FM}$ and $ W_{AFM} $ of the FM and AFM
states, for which $W_m=W_{FM},W_{AFM}$ respectively, become equal
at some field $H^{*}=2(W_{FM}-W_{AFM})/\mu $, and, if these were
the only states possible, one could expect a first order phase
transition from the AFM to the FM structure at $H=H^{*}$. However,
numerous other states are possible, and the situation is very much
more complicated.

In the intermediate region between AFM and FM, numerous more
complex structures with $0<\left\langle \mu \right\rangle <\mu $
("ferrimagnetic" structures) may occur. For these states, the
dipole-dipole interaction energy, $W_m$, is higher than $W_{AFM}$,
that for the optimal chess AFM, but the total energy $W_{m,H}
=W_m-\left\langle \mu \right\rangle H$ is reduced with increasing
field $H$. Therefore, such structures may constitute the ground
state at finite magnetic fields. To describe these structures, it
is convenient to introduce the dimensionless magnetization
$m=\left\langle \mu \right\rangle /\mu $.

In order to determine the values of $H$ that fix the lower and
upper bounds of such ferrimagnetic ground states, we have
calculated the change in dipole-dipole interaction energy $\Delta
W_m$ that occurs when the magnetic moment of a single dot is
reversed in the FM and chessboard AFM structures. A simple
analysis shows that this energy change is determined by the energy
per dot in the initial states. Reversing the magnetic moment of
one dot in the FM state requires $\Delta W_m=4W_{FM}$, and in the
chessboard AFM state $\Delta W_m=4|W_{AFM}|$. The value of
$W_{FM}$ found numerically is $4.516811\mu ^2/a^3$. Including the
magnetic field, one can show that the total energy $W_{FM}$ of the
pure FM state and that of the same state, but with one dot
reversed, coincide at the value $H=H_1$, where

\begin{equation}
H_1=2W_{FM}/\mu =9.033622\text{ }\mu /a^3.  \label{criticfield}
\end{equation}
Evidently, for $H>H_1$ the FM structure is the most favorable, but
with $H<H_1$, some magnetic moments tend to reverse. When $H<H_1$
, but close to $H_1$, the density of these reversed moments will
be very low and $ \left\langle \mu \right\rangle \approx \mu $.

The chessboard AFM state can be treated in the same way. One
obtains $W_{AFM}=-1.322943\mu ^2/a^3$ and for the corresponding
threshold field $H_0$, above which it becomes favorable to switch
a dot from antiparallel to parallel to the field direction, one
finds $H_0=2|W_{AFM}|/\mu =2.645886$ $\mu /a^3$.

Hence, intermediate phases with $0<\left\langle \mu \right\rangle
<\mu $ exist within the finite range of fields, $H_0<H<H_1$, and
$\left\langle \mu \right\rangle \rightarrow 0$ as $ H\rightarrow
H_0$, and $\left\langle \mu \right\rangle \rightarrow 1$ as
$H\rightarrow H_1$. Let us consider the nature of the ground
states in this field range. In the limit of low fields, these
states are obtained from the chessboard AFM by reversing the
magnetic moments of a small fraction of the down dot population,
leaving the remainder undisturbed. At high fields $H\approx H_1$,
the initial structure is the FM one. In both cases, one expects
the flipped dots to be dispersed as far from each other as
possible, in order to minimize their contribution to the
dipole-dipole interaction energy of the system. If it were not for
the constraints imposed by the square lattice on which all the
dots are located, one would therefore expect these flipped dots to
lie on an equilateral triangular lattice.

\begin{figure}
\includegraphics[width=3.5in]{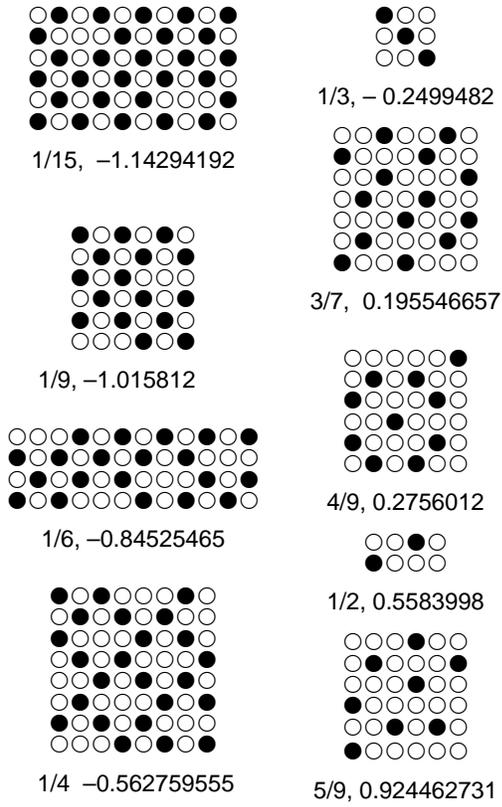}
\caption{
\label{Fig2}   Some lattice structures with intermediate values of
$m=\left\langle \mu _z\right\rangle /\mu $, $0<m\leq 5/9$, and
thought to have least dipolar energy for that value of $m$. The
values of $m$ and dipolar energy per dot of each structure (in
units of $\mu ^2/a^3$) are printed below it.}
\end{figure}

\begin{figure}
\includegraphics[width=3.5in]{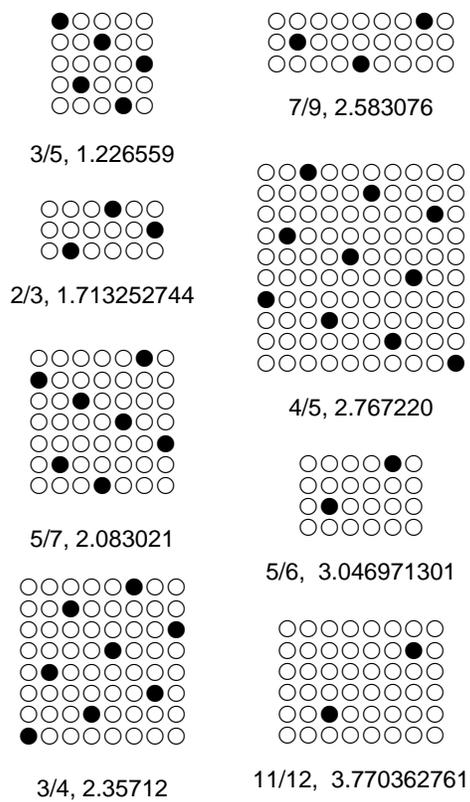}
\caption{
\label{Fig3}  As Fig.2, but for $3/5\leq m<1$.}
\end{figure}

Guided by these considerations, we have sought and found excellent
candidates for those configurations for which $W_m$ is minimal,
for a number of fixed values of $ m=\left\langle \mu \right\rangle
/\mu $ between $0$ and $1$. The elementary rectangular magnetic
cells that represent a selection of these "optimal" (we drop the
parentheses hereafter) configurations are depicted in Figs.\
\ref{Fig2} and\ \ref{Fig3}, together with their corresponding
values of $m$ and $W_m$ (with $W_m$ expressed in units of $\mu
^2/a^3$). These $W_m$ (also the values for $ W_{FM} $ and
$W_{AFM}$ given above) were evaluated numerically by summing the
contributions to the field, at each dot in the cell, of all dots
within a radius $10,000.5a$. The contribution from all more
distant dots was approximated by attributing a uniform areal
dipolar moment density $m\mu /a^2 $ to the area of the plane of
the dots outside that radius. All values of $W_m$ obtained in this
way are plotted as the points in Fig.\ \ref{Fig4}.

The sequence of configurations in Figs.\ \ref{Fig2} and\
\ref{Fig3} represent some of the (very many) stages in the
anhysteretic magnetization of the dot array from the demagnetized
AFM to the fully magnetized FM state. Owing to the stability of
the vortex state in the dots, it is evident that these states
probably cannot be accessed sequentially merely by increasing the
applied field. They represent stages in an \textit{ideal}
anhysteretic sequence, probably accessible only by thermal, or
quasi-thermal (e.g. magnetic "shaking") cooling through the Curie
temperature (or quasi-Curie temperature), under the appropriate
constant normal magnetic field $H(m)$.

It is also evident that the configurations reported here
constitute only a small sample drawn from an infinite sequence of
such optimal configurations over the range $ 0\leq m<1$: for every
rational value of $m$ in this range, there exist, in principle,
numerous different configurations, one (or possibly more) of which
must possess the lowest value of $W_m$. (Henceforward, unless the
contrary is explicit, $W_m$ will be used exclusively to refer to
this lowest energy for given $m$.)

Less evident, but at least extremely plausible, is the hypothesis
that $W_m$ increases monotonically with $m$. Consider the optimal
configurational state, in zero applied field, for any specific
reduced magnetization $m$. In both the majority "up" dot and
minority "down" dot populations, the dots are occupying the
"energetically best" locations available to them. However, by
virtue of being in the minority population, even the least
favorably located of the "down" dots is surely more favorably
located than the least favorably located "up" dot: it has more
dots of opposite sign with which to interact, and need have no dot
of the same sign for a nearest neighbour; the least favorably
located "up" dot, by contrast, is certain to have another "up" dot
alongside. How, then can it be energetically favorable to reverse
the moment of a "down" dot, thereby creating yet another "up" dot?
Indeed, this argument can be pushed a stage further: not only must
$W_m$ increase monotonically with $m$, but so must its rate of
increase, $dW_m/dm$, because the dots being reversed are selected
in order of increasing stability and new sites for reversed dots
are increasingly less favorable. It appears, however, that
$dW_m/dm$, though monotonically increasing, is discontinuous. For
example, consider a state with a simple structure like that for
$m=1/2$ in Fig.\ \ref{Fig2}. It is evident that the energy
increase on reversing one of the minority down dots is
substantially greater than the energy decrease on reversing one of
the majority up dots (taking into account the decrease in the
former energy change and increase in the latter on optimizing the
two new states). Indeed we have carried out this procedure for all
but one (that for $m=27/28$ which is very close to the FM state)
of our optimal configuration candidates, all of which support this
prediction. This aspect is further discussed in Section IV.

\section{ANALYTIC ASYMPTOTIC APPROXIMATIONS}

\label{sec:analytics}

Analytic formulae designed to approximate the dipolar energy $W_m$
in the limits $m$ $\rightarrow 0$ and $m\rightarrow \infty $ will
now be derived. These formulae are both instructive and in
remarkably good accord with the numerical values of $W_m$
calculated for specific states and represented by the points
plotted in Fig.\ \ref{Fig4}.

\subsection{Approximations near FM state}
\subsubsection {Equilateral triangular superlattice}

Consider states with $m=1-\epsilon , 0 < \epsilon  \ll 1$. In this
region, as mentioned earlier, one expects the minority down dots
to be distributed as far from each other as possible, thereby
minimizing their positive dipolar interactions with one another
and maximizing their negative interactions with the majority up
dots. For a given density of minority dots, as prescribed by
$\epsilon  $, their maximum separation is known to be ideally
accomplished when those dots form an equilateral triangular
lattice (ETL hereafter). In the present case this cannot be
precisely achieved because all the dots are constrained to lie
only at points on the fundamental square lattice of spacing $a$.
However, when $\epsilon  $ is very small, the spacing of the
minority dots $\lambda \gg a$ , and they can adopt a fair
approximation to an equilateral triangular distribution, and
indeed, as $\epsilon \rightarrow 0$, this approximation will
become very good. In developing our approximate formula, we will
therefore assume the minority dots all lie on an ETL with spacing
$ \lambda (\epsilon  )$.

In order to proceed, we need to know the interaction field and
energy for dots in an FM state on an ETL. This was calculated
numerically, in a manner similar to that used for distributions on
a square lattice described above, again summing over a circular
region, of radius $R=10000.5\lambda $, surrounding a central dot
at which the field of the others is calculated. At lattice spacing
$\lambda $, there are $(2/ \sqrt{3})/\lambda ^2$ dots per unit
area. The region outside $R$ was represented, as before, as
uniformly polarized with the mean dipole moment per unit area,
$(2/\sqrt{3})\mu /\lambda ^2$. The numerical calculation yields an
energy per dot $W_{FM\triangle }=5.517088$ $\mu ^2/a^3$. For $
\lambda $ $=a, W_{FM\triangle }$ is substantially higher than the
value $ W_{FM} = 4.516811$ $\mu ^2/a^3$ obtained for the square
lattice, but that is because the dot areal density is a factor
$2/\sqrt{3}$ higher. For the same dot areal density, $1/a^2$, we
require $\lambda =[2/\sqrt{3}]^{1/2}a$. Because of the inverse
cube interaction law, the energy per dot, at the same dot density,
$1/a^2$, is a factor $[(\sqrt{3})/2]^{3/2}$ \textit{lower} than
5.517088 $\mu ^2/a^3$. This gives $W_{FM \triangle }=4.446373 \mu
^2/a^3$ for an ETL of moment density $\mu /a^2$, about $1.56 \%$
lower than $W_{FM}$ for the square lattice, demonstrating the
small but significant energetic advantage of the former
configuration. (Indeed, the smallness of this energy difference,
for such very different configurations, is reassuring, for it
indicates that the small departures from the ideal ETL, that are
imposed by conformity with the underlying square lattice, will not
introduce any substantial error.) Analogous to $H_1=2W_{FM}/\mu
=9.033622$ $\mu /a^3,$ we will write $H_{\triangle
}=2W_{FM\triangle }/\mu=8.8927451$ $\mu /a^3$. The self-energy per
dot, of an ETL of dots, with moment $\mu $ per dot, and dot areal
density $(\epsilon /2a^2)$, is $(\epsilon  /
2)^{3/2}W_{FM\triangle }$ in the FM state. Returning to the
$m=1-\epsilon  $ system, we can regard it as the
\textit{superposition}, on the uniform square FM dot lattice, of
an (approximately) equilateral triangular system of
"\textit{double-dots}" of moment $-2\mu $ and of moment areal
density $-\epsilon \mu /a^2$, i.e. with spacing $ \lambda
=[(2/\sqrt{3})(\epsilon /2)]^{1/2}a$ appropriate to a dot areal
density $-\epsilon /2a^2$. We superpose double-dots in order, in
effect, to \textit{reverse} the moments $\mu $ of that fraction,
$\epsilon /2$, of dots on the basic square lattice that constitute
the (approximate) ETL of up dots that require to be reversed to
achieve the required overall reduced moment $m=1-\epsilon $. The
contribution to the energy of each double-dot, of moment $-2\mu $,
due to its interaction with all the other double-dots, is
$4W_{FM\triangle }(\epsilon /2)^{3/2}$. This contributes $4(\mu
H_{\triangle }/2)(\epsilon /2)^{5/2}$ to the mean dipolar energy
per dot of the overall square lattice, in this analytic asymptotic
approximation for $m\approx 1$, denoted $W_{a1}$. In addition, the
double-dots also experience the field $-H_1$ of the underlying
square FM lattice on whose dots they are superimposed.
Consequently the interaction energy of the triangular and square
lattices is $-\mu \epsilon H_1$, per square lattice dot. The
self-energy of the square FM lattice is, of course, just $W_{FM}$
$=\mu H_1/2$ per dot. Adding the three contributions gives for the
mean energy per dot of the $m=1-\epsilon $ system:

\begin{equation}
W_{a1}=\mu H_1(0.5-\epsilon )+4(\mu H_{\triangle }/2)(\epsilon
/2)^{5/2}.  \label{WA1}
\end{equation}

This formula is represented by the curve Wa1$(m)$ that extends
from $ m=0.5 $ to $m=1$ in Fig.\ \ref{Fig4}. Agreement with the
points $W$ that represent the numerical values calculated for
specific optimal structures is remarkably good, not only near
$m=1$, but over the whole of this range. One notes also that over
the whole range, the approximate values $W_{a1}\leq W_m$, the
"exact" numerical values for specific configurations. This is as
it should be, because, in the approximation, the minority dots are
located on an \textit{\ ideal }ETL whereas, in the configurations
treated numerically, the minority dots are restricted to points on
the square lattice that are close to, but not precisely at, the
ideal locations.

\subsubsection{ Square superlattice }

Examination of the specific configurations treated numerically and
illustrated in Fig.\ \ref{Fig3} reveals that, in some cases,
notably those for $m=3/5$ and $m=4/5$, the requirement that all
dots lie on the basic square lattice is so restrictive that the
minority dots are obliged to lie on a square superlattice, no
better approximation to the ideal ETL being available. It is
instructive therefore to modify the above treatment and adapt it
to a square lattice. Whereas the minority dots can only ever
approximately conform to an ETL, they can lie precisely on a
square lattice whenever the area per minority dot is $k^2+l^2$
with $k,l$  integers. Of the three contributions to the mean
energy per dot expressed in Eq. (\ref{WA1}), only the self-energy
of the double-dot lattice requires modification: in place of
$H_{\triangle }$ in that equation, we require $H_1$. Or, expressed
rather in terms of $W_{FM} =0.5\mu H_1$, we obtain for the energy
per dot of the system with minority dots on a square superlattice:

\begin{equation}
W_{\Box }=W_{FM}[1-2\epsilon +4(\epsilon /2)^{5/2}). \label{WA2}
\end{equation}
This expression yields precisely the same values as those found
directly numerically for the specific structures proposed for
$m=3/5$ and $m=4/5$ in Fig.\ \ref{Fig3} and also, for $m=0$, that
quoted above for $W_{AFM}$. We have
$W_{AFM}=W_{FM}[(1/(\sqrt{2})-1]$.

Because $H_{\triangle }$ and $H_1$ differ by only $1.56 \%$ and
the double-dot lattice self-energy term is proportional to
$\epsilon ^{5/2}$, $W_{\Box }$ differs very little from $W_{a1}$
over the range $1/2\leq m\leq 1$; the difference in the worst
case, $m=1/2$, is only $1.11\%$. We do not include a curve
representing $W_{\Box }$ in Fig.\ \ref{Fig4} because it can
scarcely be distinguished from that for $W_{a1}$. Over the range
$1/2 \leq m \leq 1$, whereas $W_{a1}$ represents a close
underestimate of $W_m , W_{\Box }$ provides a similarly close
overestimate.

\subsection{ Approximation near AFM state}

Here we consider states with positive $m$ close to $0$. The
treatment resembles that in the environs of the FM state discussed
above. We again expect the dots that depart from the chessboard
AFM structure ("up" dots this time), to be distributed as far from
each other as possible, in locations approximating an ETL. Again
we assume these "exceptional" dots all lie on an \textit{ideal}
ETL with spacing $\lambda $, and replace $\epsilon $ in the above
discussion of the triangular lattice energy by $ m $. We now
superpose the triangular lattice of "double-dots" on the AFM
lattice instead of the FM one, an important difference being that
we must, of course, place the positive double-dots only on top of
negative dots of the underlying AFM lattice, whereas all dots in
the FM lattice were equivalent and available for reversal.

Apart from the replacement of $\epsilon $ by $m$, the expression
for the self-energy of the double-dot triangular lattice is the
same: $4(\mu H_{\triangle }/2)(m/2)^{5/2}$ per dot of the overall
square lattice. However, the interaction energy of the triangular
and square lattices is now positive, $+m\mu H_0$ (as against
$-\epsilon \mu H_1$) per dot of the square lattice and the
self-energy of the square AFM lattice is $-\mu H_0/2$ per dot.
Adding the three contributions gives for the mean energy per dot
of the overall lattice of weak reduced magnetization $m$:

\begin{equation}
W_{a0}=\mu H_0(m-0.5)+4\mu (H_{\triangle }/2)(m/2)^{5/2}.
\label{WA3}
\end{equation}
Note that the two asymptotic approximations $W_{a0}$ and $W_{a1}$
happen to coincide at $m=0.5$. The expression for $W_{a0}$ is
represented by the curve $Wa0(m)$ that extends from $m=0$ to
$m=0.5$ in Fig.\ \ref{Fig4}. Again, agreement with the points $W$
that represent the numerical values calculated for specific
optimal structures is remarkably good, not only near $m=0$, but
over the whole range $0 \leq  m\leq 0.5$ . However, the agreement
is not quite as good as was the case for $W_{a1}$, very probably
owing to the additional constraint that only dots from the
down-dot population are available for reversal, whereas in the FM
case, all dots were available. This makes it somewhat harder (in
an actual configuration, but not, of course, in the analytic
approximation) to arrange the non-AFM dots close to the ideal
triangular lattice. This cannot, however be the only reason for
the lower agreement because, whereas for the approximation near
the FM state all $W_{a1} \leq W_m$, some $W_{a0}>W_m$. An
outstanding example is that for $m=1/3$ where $W_{a0}-W_m>0.01\mu
^2/a^3$. The reason for this is that, in an actual configuration,
the constraint, assumed in the analytic treatment for $m<0.5$,
that the ideal configuration is obtainable by reversing only some
negative dots in the chessboard AFM configuration and without any
further rearrangement of the structure, does not apply and, for
values of m sufficiently higher than zero, a lower energy than
$W_{a0}$ can sometimes be achieved by violating this supposed
constraint - see, for example, the simple configuration for
$m=1/3$ in Fig.\ \ref{Fig2}.

\section{ANHYSTERETIC MAGNETIZATION CURVE}

\label{sec:anhysteresys}

Consider two optimal configurations, one of reduced magnetization
$m$, the other of higher magnetization $(m+\delta m)$. Their
energies in a normal magnetic field $H$, $W_{m,H}=W_m-m\mu H$ and
$W_{(m+\delta m),H}=W_{(m+\delta m)}-(m+\delta m)\mu H$, will be
equal only in the field $ H_{m,(m+\delta m)}=(W_{(m+\delta
m)}-W_m)/(\mu \delta m).$ In the limit $ \delta m\rightarrow 0$,
we obtain the following differential expression for the
anhysteretic magnetization curve $H_m(m)$:

\begin{equation}
\mu H_m=dW_m/dm.  \label{anhysters}
\end{equation}
(Our remarks above concerning local discontinuities in $dW_m/dm$
and its monotonic increase with $m$ are clearly relevant here also
to $H_m$.)

In Fig.\ \ref{Fig5} we have plotted points representing
approximate values of $H_m(m)$ derived from the numerical values
of $W_m$ calculated for our set of candidates for optimal
configurations. They are labelled $dW/dm$, and represented by
circles. These are approximations to $H_m$ given by
$(W_{m(i+1)}-W_{m(i)})/\mu (m_{(i+1)}-m_{(i)})$ and plotted at the
values $ m=(m_{(i+1)}+m_{(i)})/2$ located midway between
successive configurations of the known set. (All fields are shown
in units of $\mu /a^3$.)

Also plotted in Fig.\ \ref{Fig5}, where they are labelled $Ha1$
and $Ha0$, are smooth curves representing $H_{a1}$ and $H_{a0}$,
the analytic approximations to $H_m$ obtained by differentiating
the asymptotic equations (\ref{WA1}) and (\ref{WA3}) and dividing
by $\mu $:

\begin{equation}
H_{a1}=-\mu ^{-1}dW_{a1}/d\epsilon  =H_1-(5/2)H_{\triangle
}(\epsilon  /2)^{3/2},  \label{firstfield}
\end{equation}
\begin{equation}
H_{a0}=\mu ^{-1}dW_{a0}/dm=H_0+(5/2)H_{\triangle }(m/2)^{3/2}.
\label{secondfield}
\end{equation}
(The corresponding equation for $H_{\Box }=-\mu ^{-1}dW_{\Box
}/d\epsilon  $ is not plotted in Fig.\ \ref{Fig5}: it is
practically indistinguishable from $H_{a1}$ over the appropriate
range $0.5 \leq  m\leq 1$.) Both Equations (\ref{firstfield}) and
(\ref{secondfield}) provide a good fit to $ dW_m/dm$ over their
appropriate data ranges, $0\leq m\leq 0.5$ for $ H_{a0}$ and $0.5
\leq  m\leq 1$ for $H_{a1}$. As one would expect for such
asymptotic approximations, the fits are best towards the limits
$m=0$ and $m=1$. Unlike the expressions (\ref{WA1}) and
(\ref{WA3}), from which they are derived, equations
(\ref{firstfield}) and (\ref{secondfield}) do not lead to
coincident values at $m=0.5$. That expressions (\ref{WA1}) and
(\ref{WA3}) should yield the same value for $W_m$ at $m=0.5$ is a
remarkable coincidence; that their slopes $dW_m/dm$ should also
agree there, is scarcely to be expected! In fact the discontinuity
between them is in remarkably good agreement with the
corresponding step in the data for $ dW_m/dm$ in the vicinity of
$m=0.5$.

\section{CONFIGURATIONAL STABILITY}

\label{sec:configure}

The fields $H_{+}(m)$ and $H_{-}(m)$ required to render
energetically favourable the reversal of a down or up dot in the
optimal configuration for any value of $m$, and the difference
between them, $H_{+}-H_{-}$, provide measures of the stability of
that configuration. Limiting examples of these fields, for the AFM
and FM states, namely $H_0\equiv H_{+}(m=0)$ and $ H_1\equiv
H_{-}(m=1)$, were discussed earlier in this article. (In fact, the
AFM state also has $H_{-}(m=0)\equiv -H_0$, so that it is stable
over a wide field range, $2H_0.$) As mentioned earlier, we have
determined numerically the extra energies $\Delta W_{m+}$ and
$\Delta W_{m-}$ required to reverse a down dot or an up dot
respectively (including local rearrangement of the resulting
configurations to minimize their energy) in almost all our
candidates for the optimal states of reduced magnetization $m$.

The values of the corresponding fields, $H_{\pm }(m)=\Delta
W_{m\pm }/2\mu $, are plotted in Fig.\ \ref{Fig5}. Some examples
illustrating the local re-organization of the dot distributions
after single-dot reversal are presented in Fig.\ \ref{Fig6}.
Intermediate values of $m$ were selected for display in this
figure because the patterns in that region are rather more complex
and varied than those near $m=0$ and $m=1$ and usually give rise
to greater ranges of field stability.

In Fig.\ \ref{Fig5} it is evident that a salient feature of the
fields $ H_{+} $ and $H_{-}$ in this intermediate range of
magnetizations is indeed the very substantial width of the gap
between them, $H_{+}-H_{-}$, that defines the field range of
stability of the pattern against single dot reversal. Towards the
limits $m=0$ and $m=1$, that gap shrinks towards zero. However,
for the structure of the \textit{pure} AFM state,
\textit{precisely} at $m=0$, the gap is at its widest: $2H_0$.
($H_{-}$ for the pure AFM state is off-scale in Fig.\ \ref{Fig5} )
Throughout the range $0<m<1$, one expects $ H_{+}>H_m>H_{-}$, but,
of course, we do not have values of $H_m$ at the values of m that
correspond precisely to the optimal configurations. However, one
observes in Fig.\ \ref{Fig5} that, for almost every configuration
(label $ i $), $H_{+}(m_{(i)})$ exceeds the next value of
$H_m\approx  (W_{m(i+1)}-W_{m(i)})/\mu (m_{(i+1)}-m_{(i)})$, with
$m_{(i+1)}> m_{(i)}$, while, in the reverse direction,
$H_{-}(m_{(i)})<H_{m(i-1)}$. This means that the field required to
reverse a \textit{single} down dot in one of our optimal
configurations, even after allowing for local rearrangement of the
pattern to minimize the increase in dipole-dipole interaction
energy, usually exceeds the field required to render energetically
favourable the reversal of the \textit{infinite} array of down
dots required to change (anhysteretically) the pattern to that of
the optimal configuration appropriate to the next higher value of
$m$. Similarly, to reverse a \textit{ single} up dot, also after
local rearrangement, usually requires a reduction of the field to
a value below that needed to favour the (anhysteretic) reversal of
the \textit{infinite} array of up dots needed to establish the
optimal configuration for the next lower value of $m$.

Examples may be helpful here. Compare the central rectangles for
the states $m=1/2(+)$ and $ m=5/9$ in Fig.\ \ref{Fig6}. The former
has the lower mean reduced moment, $13/24<5/9$, and is embedded in
an infinite array of even lower mean reduced moment, $ m=1/2$, yet
it requires a \textit{higher} field,
$H_{+}(m=1/2)=6.681>H_m(m=0.5277...)=6.589$ (in $\mu $/$a^3$
units), to equilibrate it with the $m=1/2$ configuration than does
the periodic $m=5/9$ pattern. Likewise, compare the central
rectangles for the states $m=1/2$ and $m=5/9(-)$ in Fig.\
\ref{Fig6}. Both have the same mean reduced moment of $1/2$, but
the $m=5/9(-)$ rectangle is embedded in an infinite array of
higher reduced moment, $m=5/9$, yet it requires a \textit{lower}
field, $ H_{-}(m=5/9)=5.986<H_m(m=0.5277...)=6.589$ (in $\mu
$/$a^3$ units), to equilibrate it with that $m=5/9$ configuration
than does the periodic $m=1/2$ pattern.

The reason for this \textit{prima facie} somewhat paradoxical
behaviour is that, when only a single dot is to be switched in an
optimal regular periodic configuration, only very local
rearrangement of the dot pattern can reduce the energy of the
resulting perturbed system whose overall pattern, remote from the
switched dot, must remain optimised for the original level of mean
magnetization. It follows that the sequence of points constituting
the ideal anhysteretic magnetization "curve" must correspond to a
discontinuous sequence of stable energetically optimal dot
configurations. Magnetization by monotonic variation of the
applied field alone will exhibit very substantial hysteresis and
will realize very few, if any, of the ideal magnetization
patterns.

\section{SUMMARY AND CONCLUSIONS}

Unbounded two dimensional arrays of thin circular disk-shaped
magnetic dots on a square lattice have been considered in the
presence of a magnetic field perpendicular to the dot plane. The
radii and thickness of the dots is such that a radially symmetric
vortex magnetization structure is stable; at radii outside a
relatively limited core, the magnetization lies almost in plane
and parallel to the rim. The net moment $\mu $ is due to the
vortex core and is normal to the dot plane; it is practically
unaffected by normal fields of magnitude comparable to those from
other dots.

\begin{figure}
\includegraphics[width=3.5in]{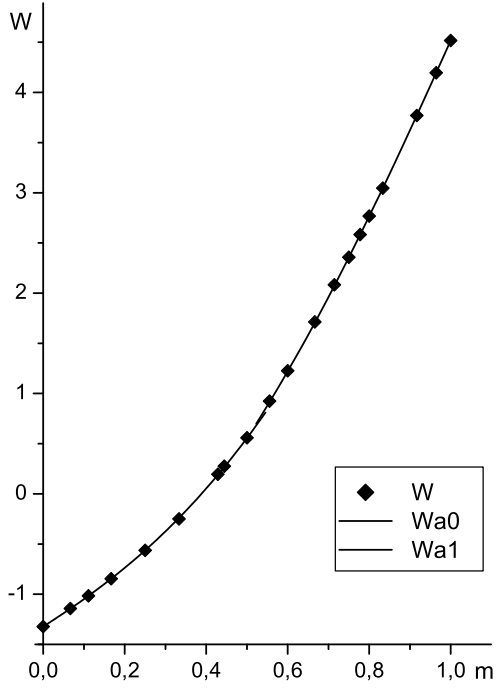}
\caption{
\label{Fig4}  Dependence of optimal lattice dipolar interaction
energy per dot, $ W_m $, on reduced lattice magnetization $m$. The
points W represent values calculated for specific structures (as
in Figs.2,3). Curves Wa1 and Wa0 represent the analytic asymptotic
formulae in Eqs.(\ref{WA1}), (\ref{WA3}). (All energies are in
units of $\mu ^2/a^3$.) }
\end{figure}

For every rational value of the reduced magnetization of the
array, $m=\left\langle \mu \right\rangle /\mu <1$, (ignoring sign)
there exist very numerous possible arrangements of up and down
dots, one (or more) of which must have minimum dipolar interaction
energy $W_m$. We have calculated that energy for a range of
excellent candidates for these ground states at various values of
$m$, in particular those of the chessboard AFM state of $m=0$ and
the uniform FM state with $m=1$. We suggest and argue that, for
the sequence of optimal configurations, $W_m$ and $dW_m/dm$
increase monotonically with $m$, and this is supported by our data
in Fig.\ \ref{Fig4}. ($dW_m/dm$ is likely to be locally
discontinuous). An analytic formula for $W_m(m)$ is derived on the
assumption that for $m\rightarrow 1$, the minority down dots are
located near points on an equilateral triangular super-lattice. A
similar formula is derived for $m\rightarrow 0$ relating to the
excess up dots. Remarkably, these formulae agree at $m=1/2$
(though their gradients differ) and fit the data for specific
states very well, especially near $m=0$ and $1$. A similar formula
involving a square super-lattice instead of the triangular
super-lattice gives results in precise agreement with the data for
specific structures with $m=0,0.6$ and $0.8$, these being those
structures in which the minority dots do indeed lie on square
super-lattices. It differs from the first asymptotic formula by
little more than $1.11\%$ over the appropriate range $0.5\leq
m\leq 1$.

\begin{figure}
\includegraphics[width=3.5in]{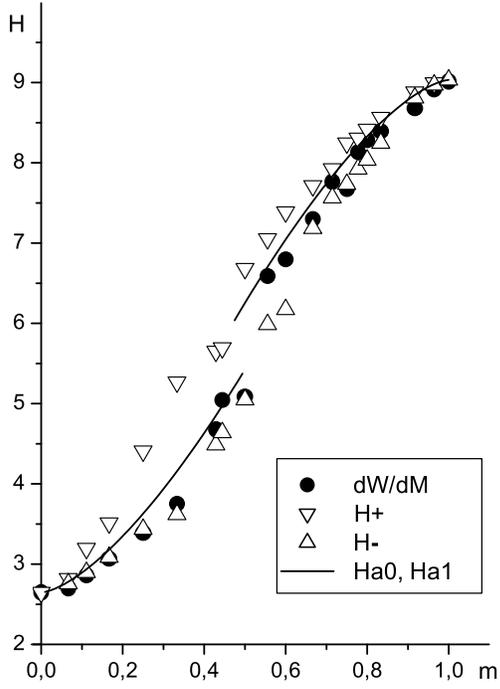}
\caption{
\label{Fig5} Anhysteretic magnetization curve for magnetic dot
lattice. Solid circles labelled dW/dM represent approximate values
of the equilibrium field $H_m(m)=\mu^{-1}dW_m/dm$ derived
numerically from the dipolar energies of the successive optimal
lattice structures studied (see text). Threshold fields for
reversing a \textit{single} dot, $H_{-}$ and $H_{+}$, are also
shown. Curves Ha1 and Ha0 follow the analytic asymptotic formulae
for $H_m(m)$ in Eqs.(\ref{firstfield}, \ref{secondfield}). (All
fields are in units of $\mu /a^3$.) }
\end{figure}

\begin{figure}
\includegraphics[width=3.5in]{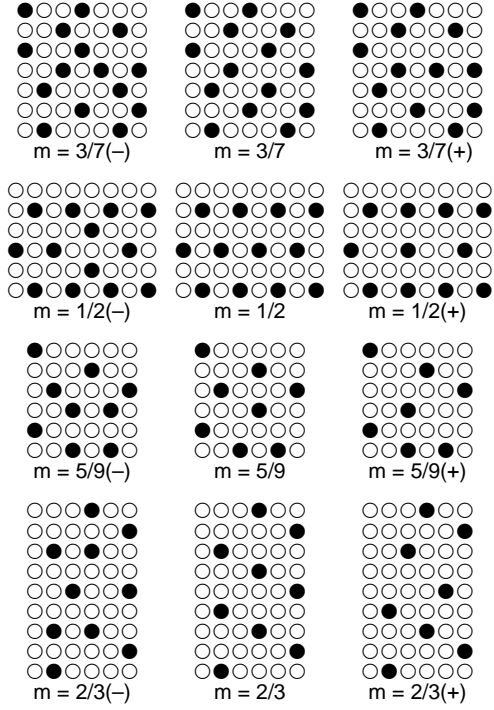}
\caption{
\label{Fig6} Examples of optimal re-ordered structures after a
single dot has been reversed in an optimal periodic lattice
configuration of specific magnetization $m$. The central column
shows the structures of individual cells in the initial periodic
rectangular superlattice. Reversing a single up dot in a single
such cell ("central" cell) and then rearranging to minimize the
energy, alters the central cell structure to that in the left
column. Similarly, the result of reversing a single down dot is
shown in the right column. None of the other cells in the
superlattice is altered. }
\end{figure}

Approximate data for the anhysteretic magnetization curve
$H_m(m)=\mu^{-1}(dW_m/dm)$ are derived from the data for specific
states and compared with the predictions of the analytic formulae
in Fig.\ \ref{Fig5}. The agreement is good, particularly near
$m=0$ and $1$. The fields predicted by the two asymptotic formulae
differ quite sharply at $m=1/2$, where they indicate a step
discontinuity in field that matches a similar step in $ H_m(m) $,
the data from the series of specific states.

The stability of the optimal configurations found was explored by
determining the minimum field $ H_{+}$ required to reverse a
single down dot and likewise the maximum field $ H_{-}$ at which a
single up dot would reverse, assuming in both cases local
reorganization of the resulting dot pattern to minimize its
energy. The gap $ H_{+}-H_{-}$ between the two fields that
indicates the stability of the configuration increases irregularly
from near zero, for $m$ just greater than zero, to a rough maximum
around $m=1/2$ and then decreases again towards zero at $m=1$. An
exception to this general trend is the very large value at $m=0$
where $H_{+}=-H_{-}=H_0$. For the optimal configurations at most
values of $m$, $H_{+}>H_{m+}$, and $H_{-}<H_{m-}$, where $H_{m+}$
and $ H_{m-}$, refer to the optimal configurations studied at
neighbouring values of magnetization, just above and just below
$m$. It follows that there exists, with increasing m, a sequence
of stable optimal configurations with energy barriers between
them. It must be stressed, therefore, that the field curve
$H_m(m)$ is strictly an ideal \textit{anhysteretic} magnetization
curve: owing to the stability of the individual dot moments, the
individual dot distributions are likewise very stable, and the
infinite sequence of energetic optimal states cannot be traced
experimentally by monotonically increasing (or decreasing) the
normal field alone. To realize any specific state it would be
necessary - but perhaps not sufficient - to cool the sample
through the Curie temperature subject to a normal magnetic field
of the appropriate strength. Another possibility would be to
destroy the stability of the other metastable phases and achieve
the phase with minimal energy by a form of magnetic shaking, e.g.
by the application of fluctuating fields of decreasing amplitude.

For magnetic storage applications, on the other hand, the
stability of the dot configurations is advantageous, indeed,
essential. Experimental observation of these dot arrays and
transitions between their states may perhaps also be useful for
the determination of the basic dot parameters, in particular the
radius of the vortex core and its magnetic moment.

Last, but not least: these results are clearly not restricted to a
dot lattice of the type considered here, but also apply directly
to any square lattice of identical dipoles that are restricted to
the two senses of normal orientation. They may also be relevant to
the description of other uniaxial dipole-coupled systems of small
particles, for example, thin films of granular magnets with easy
axial anisotropy (shape or crystallographic) with a perpendicular
easy axis and negligible exchange coupling between granules. Such
properties are characteristic for thin films prepared by
simultaneous evaporation of permalloy and silver with small enough
concentrations of permalloy.\cite{Pod'yalovski}

\begin{acknowledgements}
This work was supported in part by grant INTAS - 97 - 31311
\end{acknowledgements}

\end{document}